\documentclass[preprint,showpacs,preprintnumbers,amsmath,amssymb,11pt]{revtex4}
\usepackage{graphicx}% Include figure files
\usepackage{dcolumn}% Align table columns on decimal point
\usepackage{bm}% bold math
\begin{document}
%%%%%%%%%%%%%%%%%%%%%%%%%%%%%%%%%%%%%%%%%%%%%%%%%%%%%%%%%%%%%%%%%%%%%%%%%%%%%%%%%%
\title{Warm Inflation on the Brane}%Title of paper
%%%%%%%%%%%%%%%%%%%%%%%%%%%%%%%%%%%%%%%%%%%%%%%%%%%%%%%%%%%%%%%%%%%%%%%%%%%%%%%%%%
\author{M. Antonella Cid}
\email{mariaant@udec.cl}
\affiliation{Departamento de F\'{\i}sica, Universidad de Concepci\'{o}n, Casilla 160-C, Concepci\'{o}n, Chile}
%%%%%%%%%%%%%%%%%%%%%%%%%%%%%%%%%%%%%%%%%%%%%%%%%%%%%%%%%%%%%%%%%%%%%%%%%%%%%%%%%%
\author{Sergio del Campo}
\email{sdelcamp@ucv.cl}
\affiliation{ Instituto de F\'{\i}sica, Pontificia Universidad Cat\'{o}lica de Valpara\'{\i}so, Casilla 4059, Valpara\'{\i}so, Chile.}
%%%%%%%%%%%%%%%%%%%%%%%%%%%%%%%%%%%%%%%%%%%%%%%%%%%%%%%%%%%%%%%%%%%%%%%%%%%%%%%%%%
\author{Ram\'on Herrera}
\email{ramon.herrera@ucv.cl}
\affiliation{ Instituto de F\'{\i}sica, Pontificia Universidad Cat\'{o}lica de Valpara\'{\i}so, Casilla 4059, Valpara\'{\i}so, Chile.}
%%%%%%%%%%%%%%%%%%%%%%%%%%%%%%%%%%%%%%%%%%%%%%%%%%%%%%%%%%%%%%%%%%%%%%%%%%%%%%%%%%
\date{\today}% It is always \today, today, but any date may be explicitly specified
%%%%%%%%%%%%%%%%%%%%%%%%%%%%%%%%%%%%%%%%%%%%%%%%%%%%%%%%%%%%%%%%%%%%%%%%%%%%%%%%%%
\begin{abstract}
In this paper, warm inflationary models on a brane scenario are studied. Here we consider slow-roll inflation and high-dissipation regime in a high-energy scenario. General conditions required for these models to be realizable are derived. We describe scalar and tensor perturbations for these scenarios. Specifically we study power-law potentials considering a dissipation parameter to be a constant on the one hand and $\phi$ dependent on the other hand. We use recent astronomical observations to restrict the parameters appearing in our model.
\end{abstract}
%%%%%%%%%%%%%%%%%%%%%%%%%%%%%%%%%%%%%%%%%%%%%%%%%%%%%%%%%%%%%%%%%%%%%%%%%%%%%%%%%%
\pacs{98.80.Cq}% PACS, the Physics and Astronomy Classification Scheme.
%%%%%%%%%%%%%%%%%%%%%%%%%%%%%%%%%%%%%%%%%%%%%%%%%%%%%%%%%%%%%%%%%%%%%%%%%%%%%%%%%%
\maketitle
%%%%%%%%%%%%%%%%%%%%%%%%%%%%%%%%%%%%%%%%%%%%%%%%%%%%%%%%%%%%%%%%%%%%%%%%%%%%%%%%%%%%%%%%%%%%%%%%%%%%%%%%%%%%%%%%%%%%%%%%%%%%%%%%%%%%%%%%%%%%%%%
\section{INTRODUCTION}
%%%%%%%%%%%%%%%%%%%%%%%%%%%%%%%%%%%%%%%%%%%%%%%%%%%%%%%%%%%%%%%%%%%%%%%%%%%%%%%%%%%%%%%%%%%%%%%%%%%%%%%%%%%%%%%%%%%%%%%%%%%%%%%%%%%%%%%%%%%%%%%
It is well known that many long-standing problems of the Big Bang model, namely the horizon problem, flatness, homogeneity and the numerical density of monopoles, may find a natural explanation in the frame of the inflationary universe model \cite{Guth1981,Inflation,Inflationary}. Perhaps the most relevant feature of the inflationary universe model is that it provides a causal interpretation for the origin of the observed anisotropy in the cosmic microwave background (CMB) radiation, and also the distribution of large-scale structures \cite{WMAP1,WMAP3}. But the inflationary universe model has problems too. One of the problems in it is how to attach the observed universe to the end of the inflationary epoch; there are three possible solutions to this problem: reheating \cite{KolbandTurner}, preheating \cite{TBKLS} and warm inflation \cite{Berera1995,deOliveira1998,Berera1999,Bellini1998}. In this work we focus on the latter.

In standard inflationary universe models, the acceleration of the universe is driven by a scalar field (named inflaton) with a specific scalar potential. These kinds of model are divided into two regimes, the slow-roll and reheating ephocs.  In the slow-roll period the universe inflates and all interactions between the inflaton scalar field and any other field are typically neglected. Subsequently, a reheating period is invoked to end the period of inflation. After reheating, the universe is filled with radiation \cite{afterReheating,Inflation}, and then the universe gets connected with the Big Bang model.

Warm inflation is an alternative mechanism to have successful inflation and avoid the reheating period \cite{Berera1995}. In this kind of model, dissipative effects are important during inflation, so that radiation production occurs concurrently with the inflationary expansion. The inflaton interacts with a thermal bath via a friction term, where phenomenologically the decay of the scalar field is described by means of an interaction Lagrangian. For instance, the authors of Ref.\cite{new} take the interaction terms of the form $\frac{1}{2}\lambda^2\phi^2\chi^2$ and $g\chi\bar{\psi}\psi$, where the inflationary period presents a two-stage decay chain $\phi\rightarrow\chi\rightarrow\psi$. In this case, they reported that the damping term $\Gamma$ becomes $\frac{\lambda^3g^2\phi}{256\pi^2}$. Note that if the scalar field changes a little bit, then the friction coefficient $\Gamma$ remains almost constant. From the point of view of statistical mechanics, the interaction between quantum fields and a thermal bath could be illustrated by a general fluctuation-dissipation relation \cite{new2}. Warm inflation was criticized on the basis that the inflaton cannot decay during the slow-roll phase \cite{new3}. However, in recent years, it has been shown that the inflaton can indeed decay during the slow-roll phase (see \cite{new4} and references therein) whereby it now rests on solid theoretical grounds. On the other hand, the inclusion of the damping term into the model needs a very special scheme: for instance, in Ref.\cite{Profe} it was found that when $\Gamma$ is constant, in the slow-roll approximation a wrong value for the number of e-fold was obtained: meanwhile, in the power-law approach this problem is absent.

Warm inflation ends when the universe heats up to become radiation dominated. At this epoch the universe stops inflating and `smoothly' enters into a radiation dominated Big Bang phase \cite{Berera1995}. The matter components of the universe are created by the decay of either the remaining inflationary field or the dominant radiation field \cite{TaylorandBerera2000}.

In standard inflationary universe models the quantum fluctuations associated to the inflaton scalar field generate the density perturbations seeding the structure formation at a late time in the evolution of the universe. Instead, in warm inflation models, the density fluctuations arise from thermal rather than quantum fluctuations \cite{Berera2004,Berera1995}. These fluctuations have their origin in the hot radiation and influence the inflaton through a friction term in the equation of motion of the inflaton scalar field \cite{Berera1996}.

Several aspects of warm inflationary universe models have been studied in the past few years. The goal of the present work is to investigate warm inflationary models on brane scenarios, where the total energy density $\rho=\rho_{\phi}+\rho_{\gamma}$ is found on the brane \cite{Profe2}. The universe is filled with a self-interacting scalar field of energy density $\rho_{\phi}$ and a radiation field with energy density $\rho_{\gamma}$.

The motivation for introducing brane scenarios is the increasing interest in higher dimensional cosmological models, motivated by superstring theory, where the matter fields (related to open string modes) are confined to a lower dimensional brane, while gravity (closed string modes) can propagate in the bulk \cite{Strings}. Shiromizu et al. \cite{Shiromizu} have found the four-dimensional Einstein's equations projected onto the brane. These projections introduce some differences in the fundamental field equations, such as the Friedmann equation and, therefore, in the equations that describes the linear perturbations theory \cite{Brane}. Our aim is quantify the modifications of the warm inflation model in the brane scenario for arbitrary inflaton potentials on the brane. In order to do this we study the linear theory of cosmological perturbations for a warm inflationary model on a brane. The perturbations are expressed in term of different parameters appearing in our model: these parameters will be constrained from the WMAP three-year data \cite{WMAP3}.

In section II we describe the dynamics of our model and we establish some approximations that we use in this work. In section III we investigate the linear theory of perturbations. Here we calculate the scalar perturbations in the longitudinal gauge and also the tensor perturbations. In section IV we take a power-law potential and we investigate the high-energy and high-dissipation regime considering a power-law dissipation coefficient $\Gamma$. Finally, in section V, we give some conclusions. We have used units in which $c=\hbar=1$.
%%%%%%%%%%%%%%%%%%%%%%%%%%%%%%%%%%%%%%%%%%%%%%%%%%%%%%%%%%%%%%%%%%%%%%%%%%%%%%%%%%%%%%%%%%%%%%%%%%%%%%%%%%%%%%%%%%%%%%%%%%%%%%%
\section{WARM INFLATIONARY MODEL ON A BRANE}
%%%%%%%%%%%%%%%%%%%%%%%%%%%%%%%%%%%%%%%%%%%%%%%%%%%%%%%%%%%%%%%%%%%%%%%%%%%%%%%%%%%%%%%%%%%%%%%%%%%%%%%%%%%%%%%%%%%%%%%%%%%%%%%%%
\subsection{Brane Scenarios in Cosmology}
If Einstein's equations hold in five dimensions, with a cosmological constant term as source, and the matter fields are confined to the 3-brane, then Shiromizu et al. \cite{Shiromizu} have shown that the projected Einstein's equations onto the brane are given by
\begin{equation}
G_{\mu\nu}=-\Lambda_4g_{\mu\nu}+\left(\frac{8\pi}{M_4^2}\right)T_{\mu\nu}+\left(\frac{8\pi}{M_5^3}\right)^2\pi_{\mu\nu}-E_{\mu\nu},
\label{eq1}
\end{equation}
where $T_{\mu\nu}$ is the energy-momentum tensor on the brane, $\pi_{\mu\nu}$ is a tensor quadratic in $T_{\mu\nu}$, $E_{\mu\nu}$ is a projection of the five-dimensionl Weyl tensor describing the effect of bulk graviton degrees of freedom on brane dynamics and $M_d$ is the Planck scale in $d$ dimensions. The effective cosmological constant $\Lambda_4$ on the brane is determined by the five-dimensional cosmological constant $\Lambda$ and the 3-brane tension $\lambda$,
\begin{equation}
\Lambda_4=\frac{4\pi}{M_5^3}\left(\Lambda+\frac{4\pi}{3M_5^3}\lambda^2\right),
\label{eq2}
\end{equation}
and the four-dimensional Planck scale is given by
\begin{equation}
M_4=\sqrt{\frac{3}{4\pi}}\left(\frac{M_5^2}{\sqrt{\lambda}}\right)M_5.
\label{eq3}
\end{equation}

In a cosmological scenario in which the metric projected onto the brane is a spatially flat Friedmann-Robertson-Walker (FRW) metric, with scale factor $a(t)$, the Friedmann equation on the brane has the form \cite{Binetruy}
\begin{equation}
H^2=\frac{\Lambda_4}{3}+\left(\frac{8\pi}{3M_4^2}\right)\rho+\left(\frac{4\pi}{3M_5^3}\right)^2\rho^2+\frac{\mathcal{E}}{a^4},
\label{eq4}
\end{equation}
where $\mathcal{E}$ is an integration constant arising from $E_{\mu\nu}$, and thus transmitting bulk graviton influence onto the brane, and $\rho$ is the total energy density on the brane. During inflation, the last term in the Eq.(\ref{eq4}) will be rapidly diluted, then, we can neglect it. We will also assume that $\Lambda_4$ is negligible, at least in the early universe. With these assumptions, we can rewrite Eq.(\ref{eq4}) as
\begin{equation}
H^2=\frac{8\pi}{3M_4^2}\rho\left(1+\frac{\rho}{2\lambda}\right).
\label{eq5}
\end{equation}
We note that in the limit $\lambda\gg\rho$ (low-energy limit) we recover the standard four-dimensional result. The quadratic modification to the Friedmann equation will be dominant at high energies and moderated $\lambda$ (the limit $\rho\gg\lambda$ is named high-energy limit), but it must be sub-dominant before nucleosynthesis. Since it decays as $a^{-8}$ during the radiation era, it will rapidly become negligible thereafter. The nucleosynthesis limit implies that $\lambda\gtrsim(1\ \text{MeV})^4$ \cite{lambda} and by Eq.(\ref{eq3}) this gives
\begin{equation}
M_5\gtrsim\left(\frac{1\ \text{MeV}}{M_4}\right)^{2/3}M_4\sim 10\text{TeV},
\label{eq6}
\end{equation}
where we know that $M_4=1.2\times10^{19}\text{GeV}$ is the four dimensional effective Planck scale. A more stringent constraint may be found if the fifth dimension is infinite. In fact, by requiring that relative corrections to the Newtonian law of gravity, which are of order $M_5^6\lambda^{-2}r^{-2}$ \cite{lambda2}, be small on scales $r\geq1\text{mm}$ and using Eq. (\ref{eq3}) we have $M_5>10^5\text{TeV}$.
%%%%%%%%%%%%%%%%%%%%%%%%%%%%%%%%%%%%%%%%%%%%%%%%%%%%%%%%%%%%%%%%%%%%%%%%%%%%%%%%%%%%%%%%%%%%%%%%%%%%%%%%%%%%%%%%%%%%%%%%%%%%%%%%%
\subsection{Background Equations in a Warm Inflationary Model on a Brane}
We consider a standard scalar field in a spatially flat FRW cosmological model in a warm inflationary scenario on a brane. The dynamics of this model is described by the following set of equations:
\begin{eqnarray}
H^2=\frac{8\pi}{3M_4^2}\left(\rho_{\phi}+\rho_{\gamma}\right)\left(1+\frac{\rho_{\phi}+\rho_{\gamma}}{2\lambda}\right)=\frac{8\pi}{3M_4^2}\left(\frac{1}{2}\dot{\phi}^2+V(\phi)+\rho_{\gamma}\right)\left(1+\frac{\frac{1}{2}\dot{\phi}^2+V(\phi)+\rho_{\gamma}}{2\lambda}\right),\label{eq7}\\
\dot{\rho_{\phi}}+3H(p_{\phi}+\rho_{\phi})=-\Gamma\dot{\phi}^2\ \ \text{or equivalently,}\ \ \ddot{\phi}+3H\dot{\phi}+V^{\prime}(\phi)=-\Gamma\dot{\phi},\label{eq8}\\
\dot{\rho_{\gamma}}+4H\rho_{\gamma}=\Gamma\dot{\phi}^2,\label{eq9}
\end{eqnarray}
where $H\equiv\dot{a}/a$ is the Hubble factor, $\rho_{\gamma}$ is the energy density of the radiation field and $\Gamma$ is the dissipation coefficient, with unit $M_4$. The dissipative coefficient is responsible for the decay of the scalar field into radiation during the inflationary regime. $\Gamma$ can be assumed to be a constant or a function of the scalar field $\phi$ or the temperature $T$ or both \cite{Berera1995}. Here, we will take $\Gamma$ to be a function of $\phi$ only. In the near future we hope to study more realistic models in which $\Gamma$ not only depends on $\phi$ but also on $T$, an expression which could be derived from first principles via quantum field theory approach \cite{Moss,Bastero}.
On the other hand, $\Gamma$ must satisfy $\Gamma=f(\phi)>0$ by the second law of thermodynamics. From now on dots  and primes mean derivatives with respect to the cosmological time $t$ and the scalar field $\phi$, respectively.
For a standard inflationary model (four-dimensional general relativity) the condition for inflation is $\dot{\phi}^2<V(\phi)$, which is equivalent to having $p_{\phi}<-\left(\frac{\rho_{\phi}}{3}\right)$, where $p_{\phi}\equiv \frac{1}{2}\dot{\phi}^2-V(\phi)$ and $\rho_{\phi}\equiv \frac{1}{2}\dot{\phi}^2+V(\phi)$. This guarantees that there is acceleration, i.e. $\ddot{a}>0$.

The modified Friedmann equation lead to a stronger condition for inflation, namely
\begin{equation}
\ddot{a}>0\ \ \Rightarrow \ \  p<-\left(\frac{\lambda+2\rho}{\lambda+\rho}\right)\frac{\rho}{3},
\label{eq10}
\end{equation}
where we have used Eqs.(\ref{eq5}) and (\ref{eq8}). We can see that in the limit $\lambda\gg\rho$ we recover the standard condition for inflation, but in the limit $\rho\gg\lambda$ we find a new condition, $p<-\frac{2}{3}\rho$. In the following we shall assume that the condition expressed by Eq.(\ref{eq10}) is satisfied. During the slow-roll approximation, we assume that the energy density is dominated by the self-interacting energy density related to the scalar field, that is $\rho_{\phi}\sim V(\phi)$, and therefore, the scalar field evolution is strongly damped \cite{Chaotic}:
\begin{eqnarray}
H^2&\simeq&\frac{8\pi}{3M_4^2} V(\phi)\left(1+\frac{V(\phi)}{2\lambda}\right),\label{eq11}\\
\dot{\phi}&\simeq&-\frac{V^{\prime}(\phi)}{\Gamma+3H},\label{eq12}
\end{eqnarray}
where '$\simeq$' means equality within the slow-roll approximation. Furthermore, as usual, we will assume a quasi-stable radiation production during the warm inflation phase \cite{Profe,deOliveira}. Thus, from Eq.(\ref{eq9}) we obtain
\begin{equation}
\rho_{\gamma}\simeq\frac{\Gamma}{4H}\dot{\phi}^2= T_r^4,
\label{eq13}
\end{equation}
where $T_r$ is the temperature of the thermal bath and the last equality means that the Stefan-Boltzmann law is valid in our model. It will be useful to introduce the adimensional parameter $r$ that parametrizes the dissipation for our model:
\begin{equation}
r\equiv\frac{\Gamma}{3H}.
\label{eq14}
\end{equation}
For the high (or weak) dissipation regime we have $r\gg1$ (or $r\ll1$).

We define the slow-roll parameters as
\begin{eqnarray}
\varepsilon&\equiv&-\frac{\dot{H}}{H^2}=\frac{M_4^2}{16\pi}\left(\frac{V^{\prime}}{V}\right)^2\frac{4\lambda\left(\lambda+V\right)}{\left(2\lambda+V\right)^2}\frac{1}{\left(1+r\right)},\\
\label{eq15}
\eta&\equiv&-\frac{\ddot{H}}{\dot{H}H}=\frac{M_4^2}{8\pi}\left(\frac{V^{\prime\prime}}{V}\right)\frac{2\lambda}{\left(2\lambda+V\right)}\frac{1}{\left(1+r\right)}-\varepsilon\left(\frac{\lambda}{\lambda+V}\right)^2.
\label{eq16}
\end{eqnarray}
We see that in the limit of low energy and weak dissipation (i.e. $V\ll\lambda$ and $r\ll1$) we recover the usual slow-roll parameters for the standard inflationary universe model.

Using Eqs.(\ref{eq11}) to (\ref{eq14}) we can write the energy density of the radiation field as a function of the scalar field potential in the form of
\begin{equation}
\rho_{\gamma}\simeq\frac{M_4^2r}{32\pi\left(1+r\right)^2}\left(\frac{V^{\prime}}{V}\right)^2\frac{2\lambda V}{2\lambda+V},
\label{eq17}
\end{equation}
which, by using the definition of $\varepsilon$, becomes
\begin{equation}
\rho_{\gamma}\simeq\frac{r\varepsilon}{4\left(1+r\right)}\left(\frac{2\lambda+V}{\lambda+V}\right)V,
\label{eq18}
\end{equation}
which may be written as
\begin{equation}
\rho_{\gamma}\simeq\frac{r\varepsilon}{4\left(1+r\right)}\left(\frac{2\lambda+V}{\lambda+V}\right)\rho_{\phi},
\label{eq19}
\end{equation}
where we have taken $\rho_{\phi}\sim V$.

It is well known that inflation occurs when the condition $\varepsilon<1$ (or equivalently $\ddot{a}>0$) is fulfilled. Thus, from the latter expression we get
\begin{equation}
\rho_{\phi}>\frac{4(1+r)}{r}\left(\frac{\lambda+V}{2\lambda+V}\right)\rho_{\gamma}.
\label{eq20}
\end{equation}
Therefore, we could have a warm inflationary period on the brane when this latter inequality is satisfied. On the other hand, inflation ends when the universe heats up, at the time in which $\varepsilon\simeq1$, which implies
\begin{equation}
\rho_{\phi}\simeq\frac{4(1+r)}{r}\left(\frac{\lambda+V}{2\lambda+V}\right)\rho_{\gamma},
\label{eq21}
\end{equation}

Finally, in the following, we would like to refer to the number of e-foldings. For our model the number of e-foldings at the end of inflation is given by
\begin{equation}
N(\phi)=-\frac{8\pi}{M_4^2}\int_{\phi}^{\phi_f}\frac{V}{V^{\prime}}\left(1+\frac{V}{2\lambda}\right)(1+r)d\phi.
\label{eq22}
\end{equation}
We see that in the limit of high-energy and high-dissipation ($V\gg\lambda$ and $r\gg1$) the rate of expansion is increased respect to a standard inflationary model, while in the limit low energy and weak dissipation ($V\ll\lambda$ and $r\ll1$) the rate of expansion is the same than that of the standard inflationary model. From now on, the subscripts $i$ and $f$ are used to denote the beginning and the end of inflationary period, respectively.

%%%%%%%%%%%%%%%%%%%%%%%%%%%%%%%%%%%%%%%%%%%%%%%%%%%%%%%%%%%%%%%%%%%%%%%%%%%%%%%%%%%%%%%%%%%%%%%%%%%%%%%%%%%%%%%%%%%%%%%%%%%%%%%%%%%%%
\section{PERTURBATIONS ON THE BRANE}
%%%%%%%%%%%%%%%%%%%%%%%%%%%%%%%%%%%%%%%%%%%%%%%%%%%%%%%%%%%%%%%%%%%%%%%%%%%%%%%%%%%%%%%%%%%%%%%%%%%%%%%%%%%%%%%%%%%%%%%%%%%%%%%%%%%%%%%
In this section we will describe scalar and tensor perturbations, where the former will be determined by using the longitudinal gauge, for a warm inflationary universe model on a brane.
%%%%%%%%%%%%%%%%%%%%%%%%%%%%%%%%%%%%%%%%%%%%%%%%%%%%%%%%%%%%%%%%%%%%%%%%%%%%%%%%%%%%%%%%%%%%%%%%%%%%%%%%%%%%%%%%%%%%%%%%%%%%%%%%%%%%%%%
\subsection{Scalar Perturbations}
By using the longitudinal gauge in the perturbed FRW metric, we write
\begin{equation}
ds^2=(1+2\Phi)dt^2-a^2(t)(1-2\Psi)\delta_{ij}dx^idx^j,
\label{eq23}
\end{equation}
where $\Phi=\Phi(t,$\textbf{x}$)$ and $\Psi=\Psi(t,$\textbf{x}$)$ are the Bardeen's gauge-invariant variables \cite{Bardeen}. The spatial dependence of all perturbed quantities are of the form $e^{i\textbf{kx}}$, where $k$ is the wavenumber. The set of perturbed Einstein field equations is
\begin{eqnarray}
\dot{\Phi}+H\Phi=\frac{4\pi}{M_4^{2}}\left(  -\frac{4}{3k}\rho_{\gamma}av+\dot{\phi}\delta\phi\right)  \left(  1+\frac{1}{\lambda}\left(
\rho_{\gamma}+\frac{1}{2}\dot{\phi}^{2}+V\right)  \right), \label{eq24}\\
\ddot{\left(\delta\phi\right)}+\left(  3H+\Gamma\right)  \dot{\left(\delta\phi\right)}+\left(  \frac{k^{2}}{a^{2}}+V^{\prime\prime}
+\dot{\phi}\Gamma^{\prime}\right)  \delta\phi=4\dot{\phi}\dot{\Phi}+\left(\dot{\phi}\Gamma-2V^{\prime}\right)  \Phi,
\label{eq25}\\
\dot{\left(\delta\rho_{\gamma}\right)}+4H\delta\rho_{\gamma}+\frac{4}{3}ka\rho_{\gamma}v=4\rho_{\gamma}\dot{\Phi}+\dot{\phi}^{2}\Gamma^{\prime
}\delta\phi+\Gamma\dot{\phi}\left(  2\left(  \delta\phi\right)  ^{\cdot}-3\dot{\phi}\Phi\right), \label{eq26}
\end{eqnarray}
and
\begin{equation}
\dot{v}+\frac{\Gamma\dot{\phi}^{2}}{\rho_{\gamma}}v+\frac{k}{a}\left(\Phi+\frac{\delta\rho_{\gamma}}{4\rho_{\gamma}}+\frac{3\Gamma\dot{\phi}}{4\rho_{\gamma}}\delta\phi\right)=0, \label{eq27}
\end{equation}
where $v$ appears from the decomposition of the velocity field $\delta u_j=-\frac{iak_j}{k}ve^{i\textbf{kx}}\ (j=1,2,3)$ \cite{Bardeen} and we have omitted the subscript $k$. Eq.(\ref{eq24}) coincides with that described in Ref.\cite{Brane}, while Eqs.(\ref{eq25}), (\ref{eq26}) and (\ref{eq27}) are standard equations in the linear theory of cosmological perturbations for warm inflationary models \cite{deOliveira}.

We should note that in the case of scalar perturbations the scalar and the radiation fields are interacting. Therefore, we expect isocurvature (or entropy) perturbations to be generated, apart from of the adiabatic ones. This occurs because warm inflation can be considered as an inflationary model where two fields interact, see Refs.\cite{deOliveira,Starobinsky}.

On large scales, i.e., $k\ll aH$, we need to describe the non-decreasing adiabatic and isocurvature modes. Furthermore, if we take into account the slow-roll approximaton, the previous set of equations becomes
\begin{eqnarray}
\Phi\simeq\frac{4\pi}{M_4^{2}}\left(\frac{\dot{\phi}}{H}\right)\left(1+\frac{\Gamma}{4H}+ \frac{\Gamma^{\prime}\dot{\phi}}{48H^{2}}\right)\left(  1+\frac{V}{\lambda}\right)\delta\phi,\label{eq28}\\
\left( 3H+\Gamma\right)\dot{\left(\delta\phi\right)}+\left(V^{\prime\prime}+\dot{\phi}\Gamma^{\prime}\right)
\delta\phi\simeq\left(\dot{\phi}\Gamma-2V^{\prime}\right)\Phi,\label{eq29}\\
\frac{\delta\rho_{\gamma}}{\rho_{\gamma}}\simeq\frac{\Gamma^{\prime}}{\Gamma}\delta\phi-3\Phi, \label{eq30}
\end{eqnarray}
and
\begin{equation}
v\simeq-\frac{k}{4aH}\left(\Phi+\frac{\delta\rho_{\gamma}}{4\rho_{\gamma}}+\frac{3\Gamma\dot{\phi}}{4\rho_{\gamma}}\delta\phi\right), \label{eq31}
\end{equation}
where we have used the background equations in the same approximation.

We could solve the above set of equations by introducing Eq.(\ref{eq28}) into Eq.(\ref{eq29}), where we get
\begin{equation}
\left( 3H+\Gamma\right)\dot{\left(\delta\phi\right)}+\left(V^{\prime\prime}+\dot{\phi}\Gamma^{\prime}\right)
\delta\phi\simeq\frac{4\pi}{M_4^{2}}\left(\dot{\phi}\Gamma-2V^{\prime}\right)\left(\frac{\dot{\phi}}{H}\right)\left(1+\frac{\Gamma}{4H}+ \frac{\Gamma^{\prime}\dot{\phi}}{48H^{2}}\right)\left(  1+\frac{V}{\lambda}\right)\delta\phi.\label{eq32}
\end{equation}
Now, taking $\phi$ as an independent variable in place of $t$ and introducing an auxiliary function $\chi$ expressed by
\begin{eqnarray}
\chi=\frac{\delta\phi}{V^{\prime}}\text{exp}\left(  \int\frac{\Gamma^{\prime}}{\Gamma+3H}d\phi\right),\label{eq33}
\end{eqnarray}
we could rewrite Eq.(\ref{eq32}) as
\begin{equation}
\frac{\chi^{\prime}}{\chi}\simeq-\frac{9}{8}\frac{\left(\Gamma+2H\right)}{\left(\Gamma+3H\right)^{2}}\left(\Gamma+4H-\frac{\Gamma^{\prime}V^{\prime}}{12H\left(3H+\Gamma\right)}\right)\left(\frac{V^{\prime}}{V}\right)\left(\frac{1+\frac{V}{\lambda}}{1+\frac{V}{2\lambda}}\right),\label{eq36}
\end{equation}
where again we have used the background equations. A solution for the latter equation becomes
\begin{equation}
\chi(\phi)\simeq C\text{exp}\left(\int\left\{-\frac{9}{8}\frac{\left(\Gamma+2H\right)}{\left(\Gamma+3H\right)^{2}}\left(  \Gamma+4H-\frac{\Gamma^{\prime}V^{\prime}}{12H\left(3H+\Gamma\right)}\right) \left(\frac{V^{\prime}}{V}\right) \left(  \frac{1+\frac{V}{\lambda}}
{1+\frac{V}{2\lambda}}\right)  \right\}  d\phi\right),\label{eq37}
\end{equation}
where $C$ is an integration constant. With the help of Eq.(\ref{eq33}), we rewrite Eq.(\ref{eq37}) as
\begin{equation}
\delta\phi\simeq CV^{\prime}\text{exp}\left(-\int\left\{\frac{\Gamma^{\prime}}{3H+\Gamma}+\frac{9}{8}\frac{\left(\Gamma+2H\right)}{\left(\Gamma+3H\right)^{2}}\left(  \Gamma+4H-\frac{\Gamma^{\prime}V^{\prime}}{12H\left(3H+\Gamma\right)}\right) \left(\frac{V^{\prime}}{V}\right) \left(  \frac{1+\frac{V}{\lambda}}
{1+\frac{V}{2\lambda}}\right)\right\}d\phi\right).\label{eq38}
\end{equation}
Now, it will be useful to introduce the function $\aleph(\phi)$ as
\begin{equation}
\aleph(\phi)\equiv -\int\left\{\frac{\Gamma^{\prime}}{3H+\Gamma}+\frac{9}{8}\frac{\left(\Gamma+2H\right)}{\left(\Gamma+3H\right)^{2}}\left(  \Gamma+4H-\frac{\Gamma^{\prime}V^{\prime}}{12H\left(3H+\Gamma\right)}\right) \left(\frac{V^{\prime}}{V}\right) \left(  \frac{1+\frac{V}{\lambda}}
{1+\frac{V}{2\lambda}}\right)\right\}d\phi.
\label{eq39}
\end{equation}
With this definition, we could rewrite Eq.(\ref{eq38}) as
\begin{equation}
\delta\phi\simeq CV^{\prime}\text{exp}\left[\aleph(\phi)\right].\label{eq40}
\end{equation}

Thus, finally, the expression for the density perturbation becomes (see \cite{KolbandTurner} and \cite{LiddleandLyth})
\begin{equation}
\delta_H=\frac{2}{5}M_4^2\frac{\text{exp}\left[-\aleph(\phi)\right]}{V^{\prime}}\delta\phi.\label{eq41}
\end{equation}
We should note that, when we consider $\Gamma=0$ and the high-energy limit $(V\gg\lambda)$ (or the low-energy limit, $V\ll\lambda$), Eq.(\ref{eq41}), reduces to $\delta_H\simeq\frac{H}{\dot{\phi}}\delta\phi$, which coincides with the expression obtained in cool inflationary universe models \cite{Inflationary}.

The fluctuations of the scalar field are generated by thermal interaction with the radiation field instead of quantum fluctuations. Following the authors of Ref.\cite{TaylorandBerera2000}, we may write for the case $r\gg1$,
\begin{equation}
\delta\phi^2\simeq\frac{k_FT_r}{2\pi^2}, \label{eq45}
\end{equation}
where the wavenumber $k_F$ is defined by $k_F\equiv H\sqrt{3r}\geq H$, and its value corresponds to the freeze-out scale at which dissipation damps out the thermally excited fluctuations.

In the high-energy and high-dissipation approximation ($V\gg\lambda\  \text{and} \ r\gg1$), we find
\begin{equation}
\delta_H^2\simeq\frac{4}{25}M_4^4\frac{\text{exp}[-2\bar{\aleph}(\phi)]}{(V')^2}\delta\phi^2, \label{eq42}
\end{equation}
where $\bar{\aleph}(\phi)$ refers to $\aleph(\phi)$ in the high-energy and high-dissipation regime
\begin{equation}
\bar{\aleph}(\phi)
\simeq
-\int\left\{\frac{\Gamma^{\prime}}{3Hr}+\frac{9}{4}\left(1-\frac{\Gamma^{\prime}V^{\prime}}{12H\left(3Hr\right)^2}\right) \left(\frac{V^{\prime}}{V}\right)\right\}d\phi.
\label{eq43}
\end{equation}

At a first approximation, and during the slow-roll phase, the relation between the density matter fluctuation, $\delta\rho$, and the metric perturbation $\Phi$ is given by
\begin{equation}
\delta\rho\simeq V^{\prime}\delta\phi\simeq -\frac{M_4^{2}}{4\pi}3H^2(1+r)\left(1+\frac{\Gamma}{4H}+ \frac{\Gamma^{\prime}\dot{\phi}}{48H^{2}}\right)^{-1}\left(1+\frac{V}{\lambda}\right)^{-1}\Phi,\label{eq44}
\end{equation}
where we have used Eq.(\ref{eq28}), and the background equations in the slow-roll approximation. In the absence of dissipation, i.e. $\Gamma=0$, we found, in the low-energy limit $(V\ll\lambda)$ that $\frac{\delta\rho}{\rho}\simeq-2\Phi$; meanwhile, in the high-energy limit $(V\gg\lambda)$ we got $\frac{\delta\rho}{\rho}\simeq-\Phi$, which is the half of the previous result. Here we have taken $\rho\simeq V$. From now on we rewrite the parameter $\lambda$ in terms of $M_4$ and $M_5$ with the help of Eq.(\ref{eq3}).

If we replace Eq.(\ref{eq45}) and the definition of $k_F$ in Eq.(\ref{eq42}) we obtain
\begin{equation}
\delta_{H}^{2}\simeq\frac{4}{25}\frac{M_4^4}{2\pi^2}\text{exp}\left[-2\bar{\aleph}(\phi)\right]\frac{T_r}{\sqrt{\bar{\varepsilon} VV^{\prime2}}},  \label{eq46}
\end{equation}
where $\bar{\varepsilon}$ is the first slow-roll parameter in the high-energy and high-dissipation regime which results in being
\begin{equation}
\bar{\varepsilon}\simeq\frac{3}{4\pi}\frac{M_5^3}{\Gamma}\left(\frac{V^{\prime}}{V}\right)^2.
\label{eq47}
\end{equation}

The scalar spectral index is given by
\begin{equation}
n_s-1=\frac{d\ln{\delta_H^2}}{d\ln{k}} \label{eq48},
\end{equation}
where the interval in wavenumber is related to the number of e-foldings by the relation $d\ln{k(\phi)}=-dN(\phi)$. Using Eqs.(\ref{eq46}) and(\ref{eq48}) the spectral index becomes
\begin{equation}
n_{s}\simeq1-\left[\bar{\varepsilon}\left(\frac{V}{V^{\prime}}\left(\frac{1}{4}\frac{r^{\prime}}{r}-2\bar{\aleph}^{\prime}\left(\phi\right)\right)+\frac{1}{2}\right)-\frac{3}{2}\bar{\eta}\right], \label{eq49}
\end{equation}
where, the second slow-roll parameter $\eta$ (for $r\gg1$ and $V\gg\lambda$) is given by
\begin{equation}
\bar{\eta}\simeq\frac{3}{4\pi}\frac{M_5^3}{\Gamma}\left(\frac{V^{\prime\prime}}{V}\right). \label{eq50}
\end{equation}

One of the features of the 3-year dataset from WMAP is that it hints a significant running in th scalar spectral index expressed by the parameter $\alpha_s$ which is defined as $\alpha_s=\frac{dn_s}{d\ln{k}}$ \cite{WMAP3}. From Eq.(\ref{eq49}), this parameter becomes
\begin{equation}
\alpha_{s}\simeq-\frac{V}{V^{\prime}}\bar{\varepsilon}\left[\frac{3}{2}\bar{\eta}^{\prime}+\frac{\bar{\varepsilon}^{\prime}}{\bar{\varepsilon}}\left(
n_{s}-1-\frac{3}{2}\bar{\eta}\right)+\bar{\varepsilon}\left(\left(\frac{V}{V^{\prime}}\right)^{\prime}\left[2\bar{\aleph}^{\prime}\left(
\phi\right)-\frac{1}{4}\frac{r^{\prime}}{r}\right]+\frac{V}{V^{\prime}}\left[2\bar{\aleph}^{\prime\prime}\left(\phi\right)-\frac{1}{4}\left(\frac{r^{\prime}}
{r}\right)^{\prime}\right]\right)\right]. \label{eq51}
\end{equation}

In models with scalar fluctuations only, the marginalized value for the derivative of the spectral index results to be $\alpha_s\sim-0.05$, from WMAP only \cite{WMAP3}.
%%%%%%%%%%%%%%%%%%%%%%%%%%%%%%%%%%%%%%%%%%%%%%%%%%%%%%%%%%%%%%%%%%%%%%%%%%%%%%%%%%%%%%%%%%%%%%%%%%%%%%%%%%%%%%%%%%%%%%%%%%%%%%%%%%%%%%%%%%%%%
\subsection{Tensor Perturbations}
%%%%%%%%%%%%%%%%%%%%%%%%%%%%%%%%%%%%%%%%%%%%%%%%%%%%%%%%%%%%%%%%%%%%%%%%%%%%%%%%%%%%%%%%%%%%%%%%%%%%%%%%%%%%%%%%%%%%%%%%%%%%%%%%%%%%%%%%%%%%%
The tensor perturbations are bounded to the brane at long wavelengths \cite{lambda2}, and decoupled from the matter perturbations to first order, so that the amplitude on large scales is equal to the classical expression for the amplitude of tensor perturbations \cite{Chaotic}. On the other hand, as was found by Bhattacharya et al. \cite{Waves}, the gravitational waves generated during inflation will be amplified by the process of stimulated emission into the existing thermal distribution. This process changes the power spectrum of tensor modes by an extra temperature dependency factor $\coth\left(\frac{k}{2T}\right)$ \cite{Waves}. The spectrum in this case is given by
\begin{equation}
A_{g}^{2}=\frac{16\pi}{M_4^{2}}\left(\frac{H}{2\pi}\right)^{2}\coth\left(\frac{k}{2T}\right)\simeq\frac{64\pi}{9}\frac{V^{2}}{M_4^2M_5^6}\coth\left(
\frac{k}{2T}\right), \label{eq52}
\end{equation}
where we have considered the high-energy approximation.

In the high-energy and high dissipation regime the spectral index $n_g$ results in being
\begin{equation}
n_g=\frac{d}{d\ln{k}}\ln\left(\frac{A_g^2}{\coth\left(k/2T\right)}\right)\simeq -2\bar{\varepsilon}, \label{eq53}
\end{equation}
where we have used that $A_g\propto k^{n_g}\coth\left(\frac{k}{2T}\right)$ (see Ref.\cite{Waves}).

From Eqs.(\ref{eq46}) and (\ref{eq52}) we may calculate the tensor-scalar ratio $R$ in the high-energy and high-dissipation regime. The result is
\begin{equation}
R(k_0)=\left(\frac{A_g^2}{P_{\mathcal{R}}}\right)_{k=k_0}\simeq\left[\frac{128\pi^3}{9}\frac{V^2\sqrt{\bar{\varepsilon}VV^{\prime2}}}{M_4^6M_5^6T_r}\text{exp}\left[2\bar{\aleph}(\phi)\right]\coth\left(\frac{k}{2T}\right)\right]_{k=k_0}, \label{eq54}
\end{equation}
where we have used $P_{\mathcal{R}}=\frac{25}{4}\delta_H^2$ and $k_0$ is referred to as the pivot point \cite{Waves}.

From the combinated set of data coming from WMAP three-year data \cite{WMAP3} and the SDSS large-scale structure surveys \cite{Tegmark} it is found that an upper bound exists for $R$, which is $R(k_0=0.002\ \text{Mpc}^{-1})<0.28\ (95\% \text{CL})$, where the value $k_0$
corresponds to $l=\tau_0k_0\simeq30$, with the distance to the decoupling surface $\tau_0=14400\text{Mpc}$. SDSS measures galaxy distributions at red-shifts $a\sim 0.1$ and probes $k$ in the range $0.016\ h\ \text{Mpc}^{-1}<k<0.011\ h\ \text{Mpc}^{-1}$. The recent WMAP three-year data give the values for the scalar curvature spectrum $P_{\mathcal{R}}(k_0)\equiv \frac{25}{4}\delta_H^2(k_0)\simeq2.3\times10^{-9}$ and the scalar-tensor ratio $R(k_0)=0.095$. These values allow us to find constraints on the parameters of our model.
%%%%%%%%%%%%%%%%%%%%%%%%%%%%%%%%%%%%%%%%%%%%%%%%%%%%%%%%%%%%%%%%%%%%%%%%%%%%%%%%%%%%%%%%%%%%%%%%%%%%%%%%%%%%%%%%%%%%%%%%%%%%%%%%%%%%%%%%%%%%%
\section{$V(\phi)=V_n\phi^n$ and $\Gamma(\phi)=\Gamma_m\phi^m$ IN THE $V\gg\lambda$ AND $r\gg1$ REGIME}
%%%%%%%%%%%%%%%%%%%%%%%%%%%%%%%%%%%%%%%%%%%%%%%%%%%%%%%%%%%%%%%%%%%%%%%%%%%%%%%%%%%%%%%%%%%%%%%%%%%%%%%%%%%%%%%%%%%%%%%%%%%%%%%%%%%%%%%%%%%%%
In order to make some explicit calculations we consider a power-law scalar potential and a power-law dissipation coefficient given by $V(\phi)=V_n\phi^n$ and $\Gamma(\phi)=\Gamma_m\phi^m$ where $n$ and $m$ are integer numbers. In the following we will restrict ourselves to the high-energy and high-dissipation regime.

With  the help of Eqs. (\ref{eq11}) and (\ref{eq12}) in the high-energy and high-dissipation regime we can rewrite the scalar field and the scale factor as functions of time in the following way
\begin{eqnarray}
\phi(t)&\simeq&\phi_i\exp\left(-\frac{V_n}{\Gamma_m}nt\right),\label{eq55}\\
a(t)&\simeq& a_i\exp\left(-\frac{4\pi\Gamma_m\phi_i^n}{3n^2M_5^3}\exp\left(-\frac{V_n}{\Gamma_m}n^2t\right)\right),\label{eq56}
\end{eqnarray}
for $n-m=2$, and
\begin{eqnarray}
\phi(t)&\simeq&\left(\phi_i^{m-n+2}-\frac{V_n}{\Gamma_m}n(m-n+2)t\right)^{\frac{1}{m-n+2}},\label{eq57}\\
a(t)&\simeq& a_i\exp\left(
\frac{4\pi\Gamma_m}{3nM_5^3(m+2)}\left(\phi_i^{m+2}-\left[\phi_i^{m-n+2}-\frac{V_n}{\Gamma_m}n(m-n+2)t\right]^{\frac{m+2}{m-n+2}}\right)
\right),
\label{eq58}
\end {eqnarray}
in the case $n-m\neq2$.

From Eqs.(\ref{eq47}) and (\ref{eq50}) we find, for the slow-roll parameters,
\begin{eqnarray}
\bar{\varepsilon}&\simeq&\frac{3M_5^3}{4\pi\Gamma_m}\frac{n^2}{\phi^{m+2}},\label{58b}\\
\bar{\eta}&\simeq&\frac{3M_5^3}{4\pi\Gamma_m}\frac{n(n-1)}{\phi^{m+2}}.\label{eq58c}
\end{eqnarray}

On the other hand, if we impose $\ddot{a}(t_f)=0$ from the expression for $a(t)$ we find the time at the end of the inflationary epoch,
\begin{equation}
t_f\simeq\frac{\Gamma_m}{n^2V_n}\ln\left(\frac{4\pi\Gamma_m\phi_i^m}{3n^2M_5^3}\right),
\label{eq59}
\end{equation}
when $n-m=2$ and
\begin{equation}
t_f\simeq\frac{\Gamma_m}{n(m-n+2)V_n}\left(\phi_i^{m-n+2}-\left(\frac{3n^2M_5^3}{4\pi\Gamma_m}\right)^{\frac{m-n+2}{m+2}}\right),
\label{eq60}
\end{equation}
for $n-m\neq2$.

From Eqs.(\ref{eq55}) and (\ref{eq57}) we find that the value of the scalar field at the end of inflation becomes
\begin{equation}
\phi(t_f)\equiv\phi_f\simeq\left(\frac{3n^2M_5^3}{4\pi\Gamma_m}\right)^{\frac{1}{m+2}},
\label{eq61}
\end{equation}
in both cases, i.e., $n-m=2$ and $n-m\neq2$.

Also, for the energy density of the radiation field there is obtained
\begin{equation}
\rho_{\gamma}\simeq\frac{3M_5^3n^2}{16\pi}\frac{V_n}{\Gamma_m}\phi^{n-m+2}\simeq\frac{1}{4}\bar{\varepsilon}\rho_{\phi},
\label{eq61b}
\end{equation}
where we have taken that $\rho_{\phi}\simeq V(\phi)$.

In relation to the number of e-foldings between the time when the perturbations exit the Hubble horizon and the end of inflation we find
\begin{equation}
N\simeq{\frac{n}{m+2}}\left[\left(\frac{V_0}{V_f}\right)^{\frac{m+2}{n}}-1\right],
\label{eq62}
\end{equation}
where $V_0=V(\phi_0)$ and $\phi_0$ represent the value of the scalar field when the scale $k_0$ was leaving the horizon.

From Eq.(\ref{eq62}) we see that we can reach the appropriate number of e-foldings in order to solve the standard cosmological problems because the scalar potential is a power-law function of the scalar field and the scalar field is a function that decreases with the cosmological time.

In the case when $m=0$ it is easy to see from Eq.(\ref{eq40}) that
\begin{equation}
\delta\phi\simeq \bar{C}\frac{\dot{\phi}}{H}(r+1)^{\frac{5}{4}}\exp\left(\frac{r}{4(r+1)}\right)\simeq \bar{C}\frac{\dot{\phi}}{H}\left(1+\frac{3}{2}r+\frac{1}{4}r^2-\frac{1}{12}r^3+O(r^4)\right),
\label{eq62b}
\end{equation}
where in the last step we have expanded the expression in terms of $r$ and $\bar{C}$ is a constant proportional to $C$. We see from Eq.(\ref{eq62b}) that when $r\ll1$ we have only adiabatic modes and in consequence the entropy modes are related to high-dissipation models. When $m\neq0$ the situation is similar but the calculations are more complicated.

For completeness, from the Eq.(\ref{eq28}), the metric perturbation $\Phi$ becomes given by
\begin{equation}
\Phi\simeq-\bar{C}\frac{\dot{H}}{H^2}(r+1)^{\frac{1}{4}}\left(1+\frac{3}{4}r\right)\exp\left(\frac{r}{4(r+1)}\right)\simeq-\bar{C}\frac{\dot{H}}{H^2}\left(1+\frac{5}{4}r+\frac{1}{8}r^2-\frac{1}{48}r^3+O(r^4)\right),
\label{eq62c}
\end{equation}
where again we have expanded the expression in terms of $r$.

From Eq.(\ref{eq46}), in the high-energy and high-dissipation limits, the scalar power spectrum becomes
\begin{equation}
P_{\mathcal{R}}(k_0)\simeq\frac{M_4^4T_r}{(n\pi)^2}\sqrt{\frac{\pi\Gamma_m}{3(M_5V_0)^3}\left(\frac{V_0}{V_n}\right)^{\frac{m+4}{n}}}
\left(\frac{V_f}{V_0}\right)^{2\left(\frac{m}{n}+\frac{9}{4}\right)}
\exp\left(\frac{3m}{8(m+2)}\left[1-\left(\frac{V_f}{V_0}\right)^{\frac{m+2}{n}}\right]\right),
\label{eq63}
\end{equation}
where $P_{\mathcal{R}}$  is related to $\delta_H^2$ by $P_{\mathcal{R}}=\frac{25}{4}\delta_H^2$.

Finally, from Eq.(\ref{eq54}), in the high-energy and high-dissipation limits, the tensor-scalar ration becomes
\begin{equation}
R(k_0)\simeq\frac{64n^2\pi^3V_0^{2}}{9(M_4M_5)^6T_r}\sqrt{\frac{3(M_5V_0)^3}{\pi\Gamma_m}\left(\frac{V_n}{V_0}\right)^{\frac{m+4}{n}}}
\left(\frac{V_0}{V_f}\right)^{2\left(\frac{m}{n}+\frac{9}{4}\right)}
\exp\left(\frac{3m}{8(m+2)}\left[\left(\frac{V_f}{V_0}\right)^{\frac{m+2}{n}}-1\right]\right)
\coth\left(\frac{k_0}{2T}\right).
\label{eq64}
\end{equation}

\begin{figure}[htbp!]
    \centering
        \includegraphics[width=0.6\textwidth]{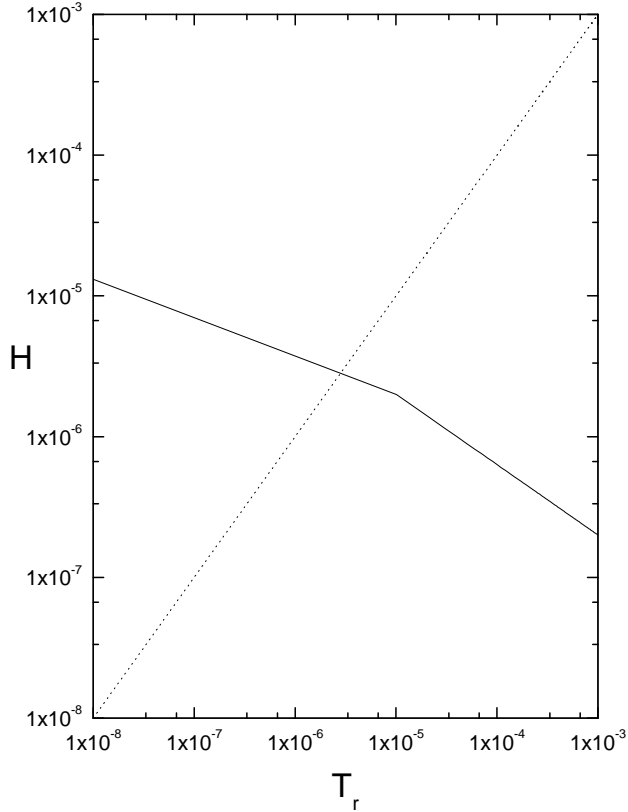}
    \caption{The Hubble parameter, $H$, as a function of the temperature $T_r$. This plot shows that the temperature needs to be greater than the value $T_r=3.42\times10^{-6}M_4$ in order to have a warm inflationary model ($T_r>H$).}
    \label{fig:HT}
\end{figure}

By resorting the WMAP three-year data \cite{WMAP3}, $P_{\mathcal{R}}(k_0)=2.3\times10^{-9}$ and $R(k_0)=0.095$. If additionally we choose $k_0=0.002\text{Mpc}^{-1}$ and $T\simeq T_r$, we can restrict the values of temperature to $T_r>3.42\times10^{-6}M_4$ (see FIG.\ref{fig:HT}) by using the necessary condition for warm inflation take place, namely, $T_r>H$ \cite{Berera1995}.

\begin{figure}[htbp!]
   \centering
     \includegraphics[width=0.6\textwidth]{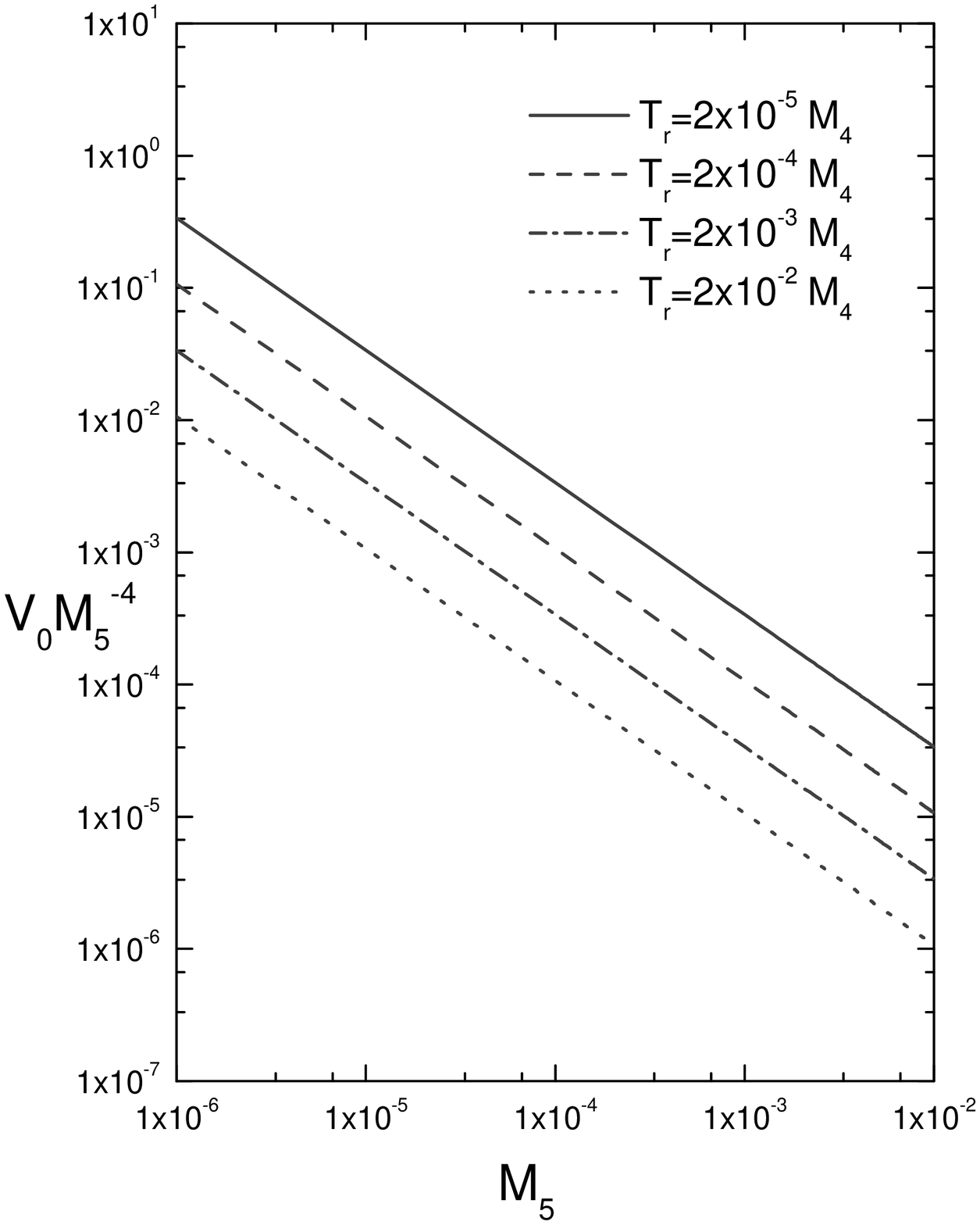}
   \caption{In this graph we show the range of $M_5$ that fulfill the requirement $V<M_5^4$ for different values of the temperature.}
   \label{fig:VM5}
\end{figure}

In the high-energy limit, i.e., $V\gg \lambda$, we need to satisfy the condition $V<M_5^4$ in order to have confined particles on the brane \cite{Binetruy}. Since the potential $V(\phi)$ is a decreasing function of time, to fulfill the condition $V(\phi)\gg\lambda$ we must have $V(\phi_f)\gg\lambda$. On the other hand, for $V(\phi)< M_5^4$ it is sufficient to have $V(\phi_0)< M_5^4$, where the period of inflation occurs between the times $t_0$ and $t_f$. Note that, for a given value of temperature, the restriction $V(\phi)< M_5^4$ allows us to establish a minimum for $M_5$  (see FIG.\ref{fig:VM5}).

In the following, we consider the restriction $V(\phi)\gg\lambda$ and we introduce a new adimensional parameter, $s\equiv\frac{V(\phi_f)}{\lambda}$, where $V(\phi_f)=V_n\phi_f^n$, with $\phi_f$ given by Eq.(\ref{eq61}).

In order to obtain some explicit values of the relevant parameters we take $r=1000$, $s=100$ and $T_r\simeq T=3.5\times10^{-6}M_4$. For these values, we found that in the case of $n=2$ and $m=0$, i.e., in the case of a chaotic potential and a constant dissipation coefficient, the spectrum is extremely near to the Harrison-Zel'dovich spectrum ($n_s=0.999$) and the running spectral index is several orders of magnitude smaller than the value given by WMAP observations \cite{WMAP3}. The values for $n_s$ and $\alpha_s$ are similar when $m=2$, which corresponds to a dissipation coefficient $\Gamma$ proportional to $\phi^2$. Additionally, in the case of a chaotic potential ($n=2$) the value of $n_s$ is extremely near to the Harrison-Zel'dovich spectrum yet, when we take any other value for the temperature (in the allowed range), or when we increase the values of $r$ or $s$.

For $T_r\simeq T=3.5\times10^{-6}M_4$, $r=1000$, $s=100$, $n=4$ and $m=0$ we have found that the value of $n_s$ is a bit different from that of the Harrison-Zel'dovich spectrum ($n_s=0.997$). Also, the value of the running spectral index is just one order of magnitud smaller than the value given by WMAP three year data ($\alpha_s=-4.60\times10^{-3}$). The other relevant parameters take the following values in this case:  $M_5=2.39\times10^{-5}M_4$, $\Gamma(\phi_0)=1.01\times10^{-2}M_4$, $V(\phi_0)=1.10\times10^{-20}M_4^4$ and $\phi_0=8.99\times10^{-5}M_4$. When $m=2$ or $m=4$ and if we consider the slow-roll approximation, we were unable to find any real solution.

Finally, the most interesting case is found when $n=6$ and $m=0$, with the same values of $T_r$, $r$ and $s$ considered in the previous case. Here, the parameters result in being $n_s=0.966$ and $\alpha_s=-6.70\times10^{-2}$, which are very closed to the values expressed by WMAP observation \cite{WMAP3}. The other parameters take the following values: $M_5=3.09\times10^{-5}M_4$, $\Gamma(\phi_0)=1.01\times10^{-2}M_4$, $V(\phi_0)=2.39\times10^{-20}M_4^4$ and $\phi_0=5.12\times10^{-5}M_4$.

%%%%%%%%%%%%%%%%%%%%%%%%%%%%%%%%%%%%%%%%%%%%%%%%%%%%%%%%%%%%%%%%%%%%%%%%%%%%%%%%%%%%%%%%%%%%%%%%%%%%%%%%%%%%%%%%%%%%%%%%%%%%%%%%%%%%%%%%%%%%
\section{Conclusions}

In this paper we have considered a warm inflationary scenario on a brane. We have restricted ourselves to a high-dissipation and high-energy regime. In the slow-roll approximation we have found a general relationship between radiation and scalar field densities; see Eq.(\ref{eq21}).

In relation to the perturbations we have considered that the there does no exist any interaction between the brane and the bulk, i.e., we have neglected back-reaction due to metric perturbations in the fifth dimension. We note that a full investigation is required to discover when back-reaction will have a significant effect in the perturbations. With the above-mentioned restriction we have obtained explicitly the contributions of the adiabatic and entropy modes. We have shown that the dissipation parameter plays a crucial role in producing the entropy mode (see Eq.(\ref{eq62b})). A general relation for the density perturbations is given in Eq.(\ref{eq41}). The tensor pertubations are generated via stimulated emission into the existing thermal background (see Eq.(\ref{eq52})) and the tensor-scalar ratio is modified by a temperature-dependent factor.

We have studied a power-law potential for different dependence of the dissipation coefficient $\Gamma$. From the normalization of the WMAP three-year data, the potential becomes of the order of $V(\phi_0)\sim10^{-20}M_4^4$ when it leaves the horizon at the scale of $k_0=0.002\text{Mpc}^{-1}$. As in the situation studied in Ref.\cite{Chaotic}, the value of the potential depends on $M_5$. Here we have considered $M_5\sim10^{-5}M_4$. To fulfill the approximations in our model we have restricted the range of the parameters in which warm inflation on a brane can occur (see FIGS. \ref{fig:HT}, \ref{fig:VM5}). In order to show some explicit results we have chosen the following set of values for the parameters, $T_r\simeq T=3.5\times10^{-6}M_4$, $r=1000$ and $s=100$. First, we have considered a chaotic potential and a constant dissipation coefficient; for this case we have found that the spectrum is driven toward an scale-invariant spectrum and the running of the spectral index is extremely small. The situation was similar when we considered a variable dissipation coefficient. In the case of a power-law potential with $n=4$ and a constant dissipation coefficient we have found that the value to $n_s$ is smaller than the previous case but out of the range given by WMAP three-year data. On the other hand, the running of the spectral index is closer to the value given by WMAP three-year data but one order of magnitud smaller.

Finally the best situation is found when we consider a constant dissipation coefficient and a $n=6$ potential. In this case both parameters, $n_s$ and $\alpha_s$, are in the ranges specified by WMAP.

It is necessary to note that with our approximations we have found that for the case in which $n=6$ and $m=0$ is favored in the light of the recent results reported by WMAP three-year data. On the other hand, we have considered some approximations; as a result, we were not able to find any real solution in the allowed range of parameters. For example, in the slow-roll approximation, in the case $n=4$ and, when we try to consider a variable coefficient of dissipation $\Gamma$, we did not find any real solution. However, we think that we could find a solution if in place of using this approximation we assume a power-law for the scale factor. We intend to return to this point (and others) in the near future.

%%%%%%%%%%%%%%%%%%%%%%%%%%%%%%%%%%%%%%%%%%%%%%%%%%%%%%%%%%%%%%%%%%%%%%%%%%%%%%%%%%%%%%%%%%%%%%%%%%%%%%%%%%%%%%%%%%%%%%%%%%%%%%%%%%%%%%%%%%%%%%%%%
\begin{acknowledgments}
M. A. C. was supported by CONICYT through the grant N$^0$21050599. S. d. C. was supported by COMISION NACIONAL DE CIENCIAS Y TECNOLOGIA through FONDECYT grants N$^0$1040624, N$^0$1051086 and N$^0$1070306. Also, from UCV-DGIP N$^0$123.787. R.H. was supported by the ``Programa Bicentenario de Ciencia y Tecnolog\'ia'' through the grant ``Inserci\'on de Investigadores Postdoctorales en la Academia'' N$^0$PSD/06.
\end{acknowledgments}

%%%%%%%%%%%%%%%%%%%%%%%%%%%%%%%%%%%%%%%%%%%%%%%%%%%%%%%%%%%%%%%%%%%%%%%%%%%%%%%%%%%%%%%%%%%%%%%%%%%%%%%%%%%%%%%%%%%%%%%%%%%%%%%%%%%%%%%%%%%%%%%%%

%%%%%%%%%%%%%%%%%%%%%%%%%%%%%%%%%%%%%%%%%%%%%%%%%%%%%%%%%%%%%%%%%%%%%%%%%%%%%%%%%%%%%%%%%%%%%%%%%%%%%%%%%%%%%%%%%%%%%%%%%%%%%%%%%%%%%%%%%%%%%%%%%%%%%
\end{document}